\documentclass[10pt,a4paper]{article}
\usepackage{amsmath,amssymb}
\usepackage{graphicx}
\usepackage{subfigure}
\usepackage{fancyhdr}
\usepackage[textwidth=14cm,textheight=22cm]{geometry}
\begin{document}
\title{\large{Super Guassian Distribution of Laser Produced Plasmas}}
\author{\large{\textbf{Kaifeng Chen}}}
\date{}
\maketitle
\clearpage
\tableofcontents
\clearpage
\section{\Large{\textbf{Introduction}}}
It is well known that the non-Maxwellian electron velocity distributions, resulting from the nonclassical drive and transport under laser-fusion relevant conditions,may significantly affect the dispersion and damping of the different waves present in the plasma. This in turn may alter the thresholds and gains of the parametric instabilities,involving the plasma wave and the ion acoustic wave,respectively.\\[0.1cm]
A systematic characterization of the electron velocity distribution,under the wide range of experimentally relevant conditions,is therefore essential in this context.Electron distributions have been computed and commented in many previous publications.But,either because they considered idealistic profiles,or considered only a small number of simulation cases,these studies did not attempt to characterize the non-Maxwellian distribution by proposing,for instance,a functional form as an appropriate fit.\\[0.1cm]
A thorough study of the non-Maxwellian electron velocity distribution functions was performed by Mattee \emph{el.al.},neglecting spatial transport. In this study,the competition between inverse Bremsstrahlung, which pushes the distribution towards a super-Guassian of the form$f\sim exp[-(v/u)^5]$,and the thermalization through self-collisions was considered.It was shown,that the distribution is well represented,at all energies,by so so-called Dum-Langdon-Matte(DLM)-type function $f_{DLM}\sim exp[-(v/u)^n]$,where the parameter $n$ takes intermediate values between $n=2$(a Maxwellian),in the absence of drive, and $n=5$,when inverse Bremsstrahlung dominates.As a practical result, a relation $n=n(\alpha)$ was established between DLM exponent $n$ and the relevant parameter $\alpha=Z(v_{os}/v_{th})^2$,measuring the relative importance of inverse Bresstrahlung compared to self-collisions. Here,$Z$ is the degree of ionization,$v_{os}$ is the quiver velocity and $v_{th}$ the thermal velocity.\\[0.1cm]
\section{\Large{\textbf{Physical Model}}}
The transport of electrons is described by the corresponding Fokker-Planck equation:
\begin{equation}
 \frac{\partial f}{\partial t}+v\cdot\frac{\partial f}{\partial x}+\frac{(-eE)}{m}E\cdot\frac{\partial f}{\partial v}=-\{C_{ee}[f,f]+C_{ei}f\}
\end{equation}
where E stands for the self-consistent electrostatic field ensuring quasineutrality,$C_{ee}$ stands for the Landau electron-electron collision operator, and $C_{ei}$ for the Lorentz electron-ion collision operator. As for the simulations, the ions are kept fixed.\\[0.1cm]
For solving the Fokker-Planck equation, a high ionic charge $Z$ is assumed, such that the electron-ion mean free path $\lambda_{ei}$ is small compared to any characteristic scale length of the system. So the distribution function can be a sum of an expansion series.
\begin{equation}
 f(x,v,t)=\sum_{l=0}^{\infty}f_l(x,v,t)P_l(\mu)
\end{equation}
where $\mu=v_x/v=cos\theta$ is the direction cosine. In the diffusion approximation.\\[0.1cm]
A factor $\alpha=Z(v_{os}/v_{th})^2$ is usually introduced to describe the relative strength of these two processes. When $\alpha$ is small, the thermalization process is dominant so that electron have a distribution close to a Maxwellian. If the heating process is dominant,i.e.$\alpha$ ids significantly larger than 1, the electron distribution should deviate from a Maxwellian and approach to a super-Guassian distribution,
\begin{equation}
 f_{DLM}(v,t)=A_n\frac{N}{v_{th}^3}exp\big[-B_n(\frac{v}{v_{th}})^n\big]
\end{equation}
where the coefficients $A_n$ and $B_n$, ensuring the definition of density($N=4\pi\int_0^{\infty}v^2dvf$) and temperature($3NT/m=4\pi\int_0^{\infty}v^4dvf$), are given in terms of the gamma function $\Gamma$ by
\begin{align}
 A_n&=\frac{1}{4\pi}\frac{n}{\Gamma(3/n)}\Big[\frac{1}{3}\frac{\Gamma(5/n)}{\Gamma(3/n)}\Big]^{3/2}\\
 B_n&=\Big[\frac{1}{3}\frac{\Gamma(5/n)}{\Gamma(3/n)}\Big]^{n/2}
\end{align}
Here $m$ is a number depending on the factor $\alpha$,
\begin{equation}
 m(\alpha)=2+3/(1+1.66/\alpha^{0.724})
\end{equation}
and $v_m$ is given by
\begin{equation}
 v_m^2=3\frac{k_BT}{m_e}\frac{\Gamma(3/m)}{\Gamma(5/m)}
\end{equation}
In this work, the data is generated by Doc. Bin Zhao's Fokker-Planck code in Epperlein's scheme. And these data contain information about the distribution function of $(x,v,t)$,temperature of each point and density of each point. All these data are fitted by using the formula proposed above.
\section{\Large{\textbf{Simulated Results}}}
\begin{figure}[!h]
\centering
\includegraphics[height=10cm,width=14cm]{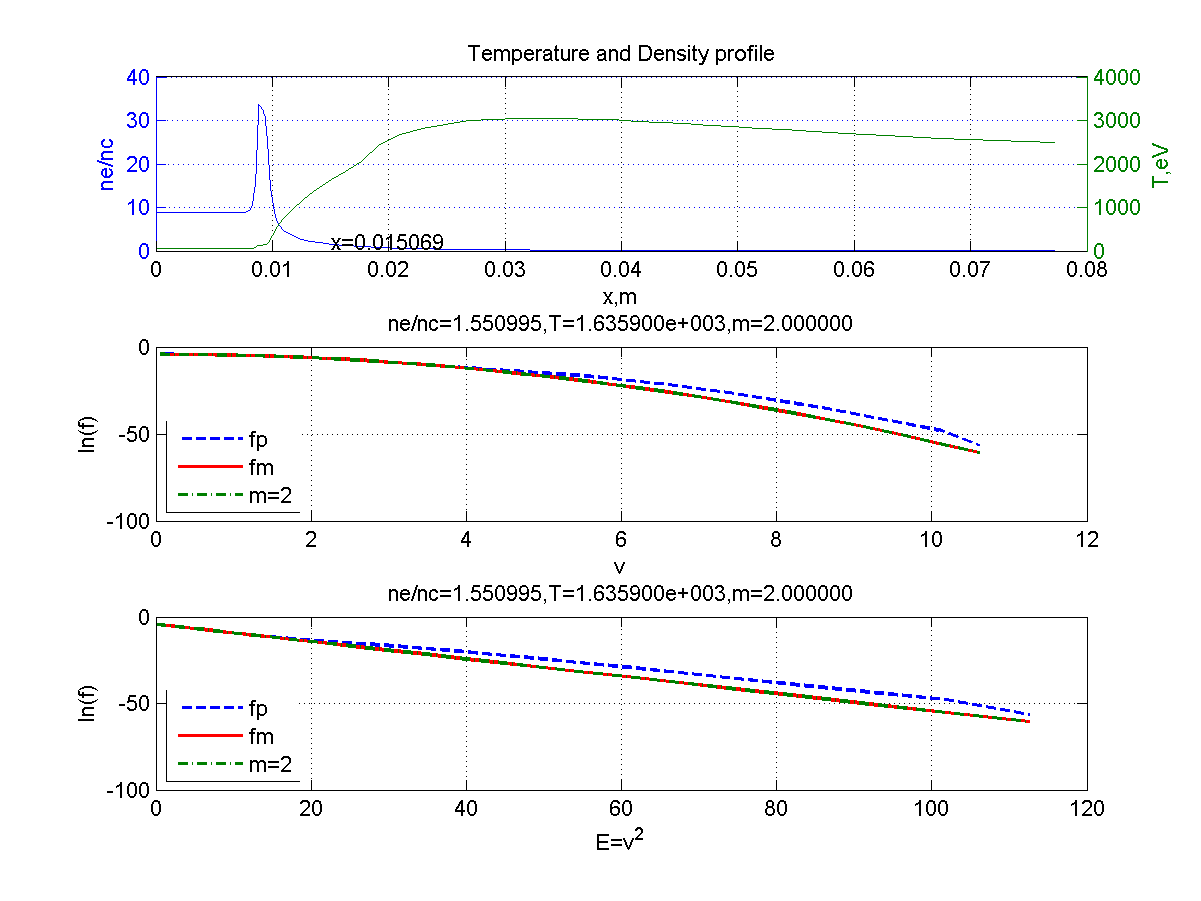}
\caption{x=75}
\end{figure}
Figure 1 shows the density,temperature and distribution function at the $75th$ lattice.As to the distribution function,the x axis has been transformed to $v^2$,which represents the energy distribution, and the y axis denote the logarithm of f. Thus for Maxwellian distribution function like $f=C_1exp(-C_2v^2)$, it will be a straight line and the slope will be 2(represented in the green line). The blue line is plotted by using DLM fitting function and the red line denotes the results calculated by Fokker-Planck equation.Some parameters are listed above each sub-picture like m,the ratio of $n_e$ to $n_c$\footnote{The critical density.},and the temperature($K$).\\[0.1cm]
In this picture, the ratio is 1.55,which means that the density is much beyond the critical density. And for this region of this high ratio,$m=2$ and the Fokker-Planck calculated data are little deviated from the DLM theory.\\[0.1cm]
Figure 2 presents the lattice of $x=76$, the ratio becomes nearly 1.27 and the three lines: pure Maxwell distribution,super-Guassian distribution function and the distribution function from Fokker-Planck equation.In this position,these three lines are almost the same because $m=2.09$,which is very close to 2.
\begin{figure}[!h]
\centering
\includegraphics[height=10cm,width=14cm]{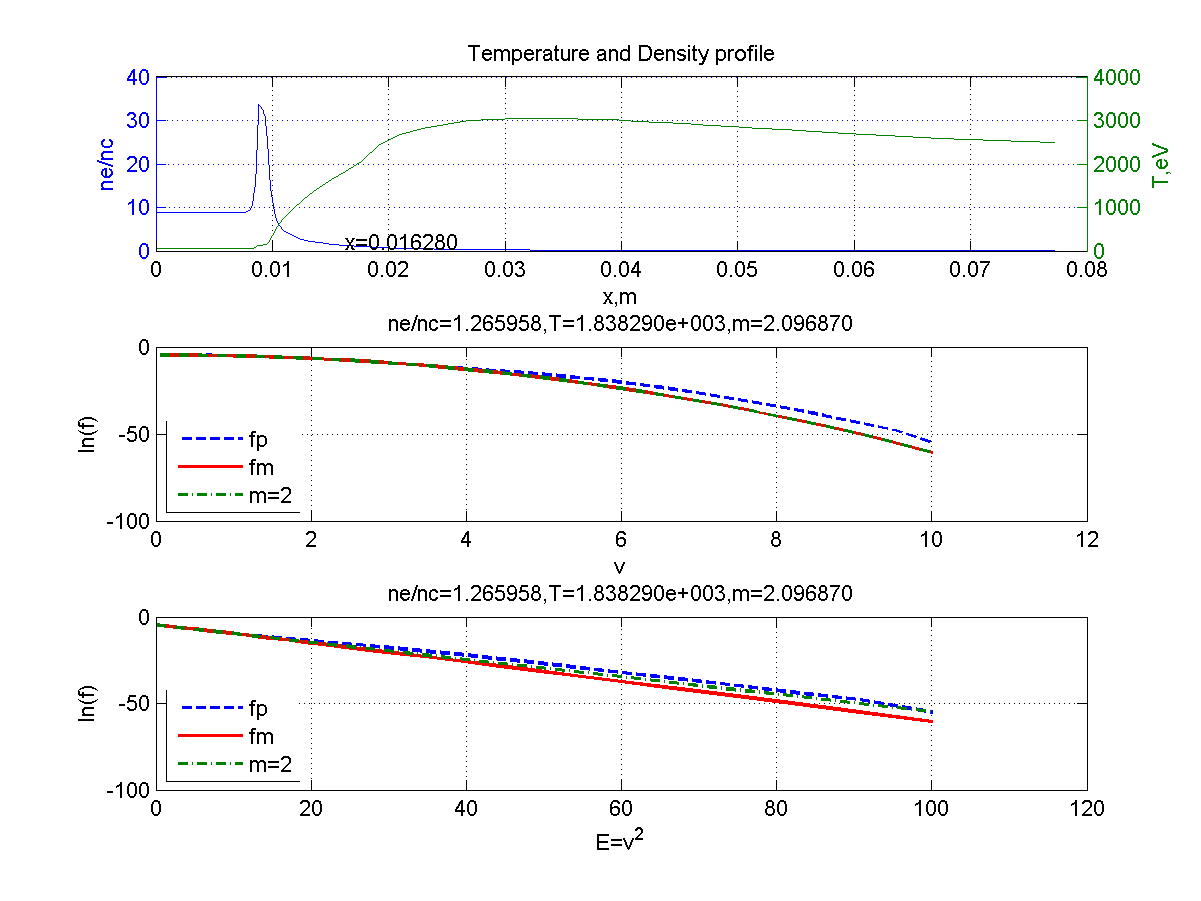}
\caption{x=76}
\end{figure}\\[0.1cm]
Figure 3 shows some slight changes when the position becomes $x=77$. In this case, the distribution calculated by Fokker-Planck equation is less than the super-Guassian fitting function at high velocity region. More over, these two functions are always higher than the Maxwellian distribution function. This phenomenon agrees well with the non-local transport theory that in high velocity region of high density heat flux is actually lower than the classical one, thus the percentage of high velocity particles will become larger.\\[0.1cm]
Under this condition,$m=2.11$ and the ratio of $n_e/n_c$ is nearly 1.00.
\begin{figure}[!t]
\centering
\includegraphics[height=9.5cm,width=14cm]{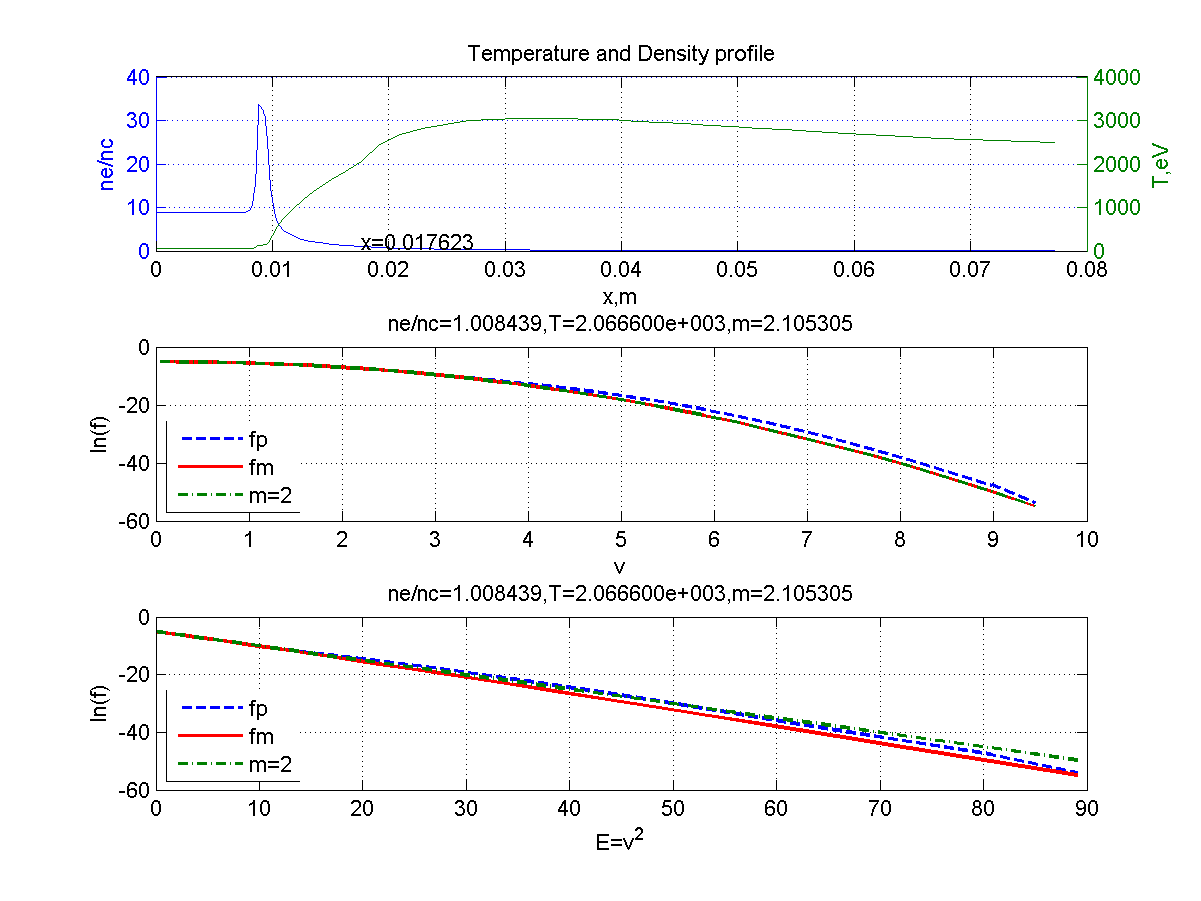}
\caption{x=77}
\end{figure}
\begin{figure}[!h]
\includegraphics[height=9.5cm,width=14cm]{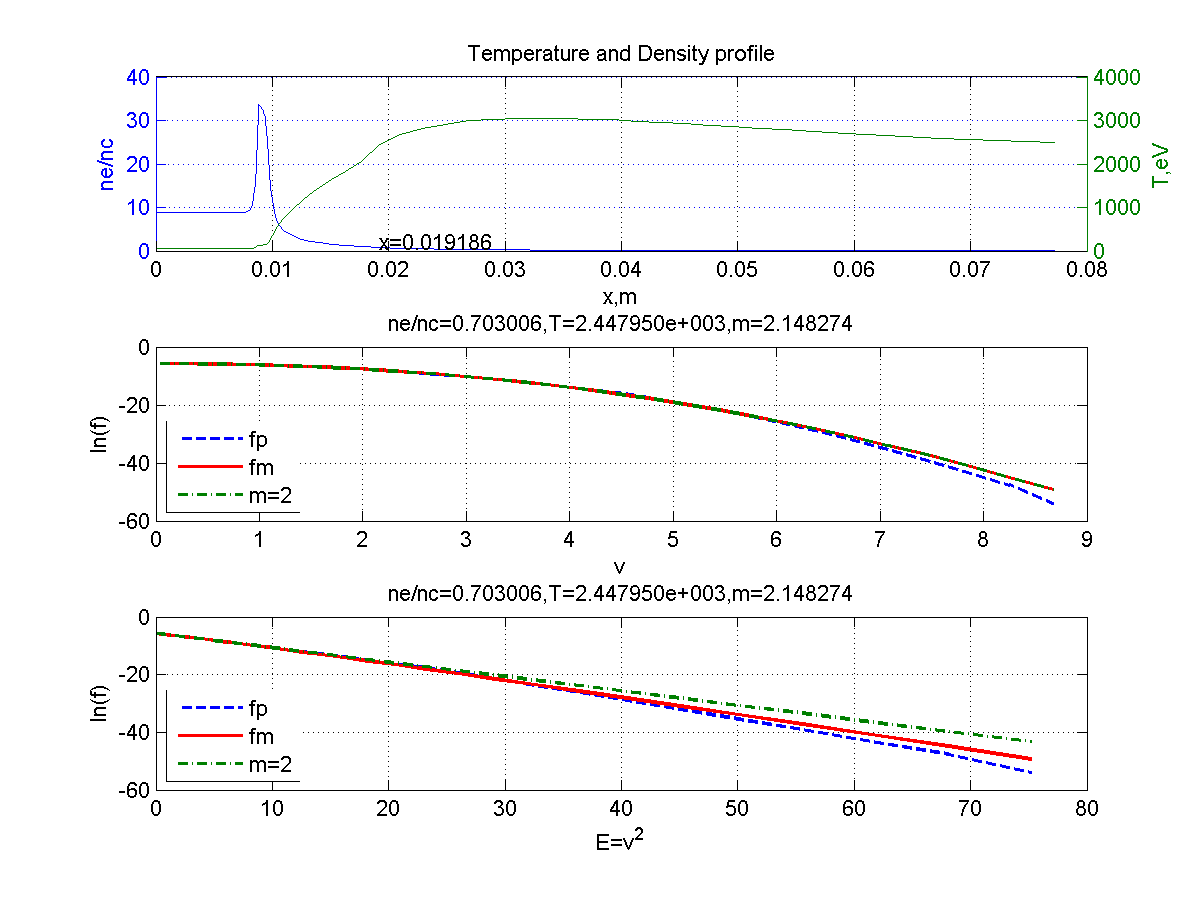}
\caption{x=78}
\end{figure}
The Figure 4 shows the results when $n_e<n_c$, which means that this position along x axis is the point away from the over density region. In this region,the Maxwellian distribution function becomes the largest one and the DLM function and the Fokker-Planck calculated result are nearly the same. More to mention,from about $v=3v_{th}$ the DLM and Fokker-Planck begin to differ from each other.This phenomenon proves the validity of Brunner's fitting function in reference 1. The value of $m$ now is 2.15.
\section{\Large{\textbf{Conclusion}}}
In summary, in the case of spacial transport and inverse Bremsstrahlung, some simulations by using Fokker-Planck equation are shown. Once the system has reached a quasisteady state, the electron velocity distributions within the laser beam are well represented by a DLM-type function for the bulk. By considering the ratio of $n_e$ and $n_c$,we can better assume the distribution functions.In the high density region, Maxwellian distribution can be used like a classical model. However, when the density goes below the critical density, DLM function should be used to simulate the non-local transport. Some evidence has shown that when $v$ is close to $3v_{th}$, there will be some deviation of the DLM-type function and Fokker-Planck calculated data. Future work will contain other cases and continuing to explore better fitting functions in the case of inverse Bremsstrahlung.
\section{\Large{\textbf{References}}}
1.S.Brunner and E.Valeo,Phy.Plasma,2002,Volume 9,Number 3.\\[0.1cm]
2.J.P.Matte,M.Lamoureux,C.Moller,R.Y.Yin,J.Delettrez,J.Virmout,and T.W.Johnston,Plasma

Phys.Controlled Fusion 30,1665(1998)\\[0.1cm]
3.J.R.Albritton,Phys.Rev.Let,1983,Volume 50,Number 26.\\[0.1cm]
4.J.P.Matte,J.Virmout Phys.Rev.Let,1982,Volume 49,Number 26.
\end{document}